%% file: main.tex
\documentclass[12 pt]{article}
\usepackage{bbm}
\usepackage{natbib}
\usepackage{amssymb}
\usepackage{amsmath}
\usepackage{amsthm}
\usepackage{setspace}
\usepackage{palatino} %use the palatino font (just cooler)
\usepackage{mathpazo} %make your math match the palatino font

\usepackage{hyperref}
\usepackage{graphicx}
\usepackage{subcaption}
\usepackage{enumitem}
\usepackage{verbatim}
\usepackage{sectsty}
\usepackage[svgnames]{xcolor}
\usepackage{subfiles}
\usepackage[capitalize,nameinlink]{cleveref}
\usepackage[colorinlistoftodos,textsize=tiny]{todonotes}
\usepackage{threeparttable}
\usepackage{xcolor}
\usepackage{changepage}
\usepackage{pgf}
\usepackage{tikz}
\usepackage{xargs}
\usepackage[font={footnotesize,it},justification=justified]{caption}
\usepackage[normalem]{ulem}
\usepackage{authblk}
\usepackage{lscape}
\usepackage{MnSymbol,wasysym}
\usepackage{booktabs}
\usepackage{soul}
\usepackage{tabularx}
\usepackage{subcaption}
\usepackage{scrextend}
\usepackage{makecell}
\usepackage{floatrow}

\hypersetup{
    pdftitle={TBA},
    pdfauthor={Ritwik Banerjee},
    colorlinks=true,
    citecolor=blue,
    bookmarksnumbered=true,
    urlcolor=blue,
    linkcolor=red
}
 \theoremstyle{definition}
 
 \theoremstyle{remark}

\theoremstyle{plain}

\newtheorem{result}{Result}

\renewcommand{\epsilon}{\varepsilon}
\usetikzlibrary{shapes,arrows,automata,fit}

\tikzstyle{info}=[circle,thick,draw=black,fill=black!25,minimum size=4mm]
\tikzstyle{uninfo}=[circle,thick,draw=black,fill=white,minimum size=4mm]
\tikzstyle{inforecog}=[circle,line width=1mm,draw=black!50,fill=black!25,minimum size=4mm]
\tikzstyle{uninforecog}=[circle,line width=1mm,draw=black!50,fill=white,minimum size=4mm]

\tikzstyle{traded}=[draw, line width=1mm]
\tikzstyle{recog}=[draw=black!50, line width=1mm]

\sectionfont{\color{DarkRed}}
\subsectionfont{\color{DarkRed}}
\subsubsectionfont{\color{DarkRed}}
\crefname{assumption}{assumption}{assumptions}
\newcommand{\appendixref}[1]{\hyperref[#1]{Appendix}}
\newcommand{\suppappendixref}[1]{\hyperref[#1]{Supplementary Appendix}}

\allowdisplaybreaks  %to split aligned equations in two pages
\usepackage{pdfpages}
\usepackage[margin=0.8in]{geometry}
\usepackage{pdflscape}
\usepackage{atbegshi}% http://ctan.org/pkg/atbegshi
\AtBeginDocument{\AtBeginShipoutNext{\AtBeginShipoutDiscard}}

\begin{document}
\begin{titlepage}
\vspace{-1cm}
\title{Using social recognition to address the gender difference in volunteering for low-promotability tasks}
\thanks{\textit{Acknowledgements}: This research has primarily been funded by the IIM Bangalore. We thank Lise Vesterlund, Sera Linardi, Maria Recalde, Anya Samek, Lata Gangadharan and seminar participants at University of Pittsburgh and ESA Global Meetings for their helpful suggestions and comments. We thank Partha Chatterjee and Sayar Mitra for helping us with conducting the experiment. We gratefully acknowledge the research assistance from Chacko Babu and Arpit Sachan. Banerjee can be reached at \href{mailto:ritwikbanerjee@iimb.ac.in}{ritwikbanerjee@iimb.ac.in} and Mustafi at \href{mailto:priyoma.mustafi@pitt.edu}{priyoma.mustafi@pitt.edu}.}

%\author{}
%\begin{comment}
\author[1]{Ritwik Banerjee}
\author[2]{Priyoma Mustafi}
\affil[1]{Indian Institute of Management Bangalore, India}
\affil[2]{University of Pittsburgh, USA}
%\end{comment}

\date{}

\clearpage\maketitle
\thispagestyle{empty}
\vspace{-1cm}
\begin{abstract}
\noindent
\small
Research shows that women volunteer significantly more for tasks that people prefer others to complete. Such tasks carry little monetary incentives because of their very nature. We use a modified version of the volunteer’s dilemma game to examine if non-monetary interventions, particularly, social recognition can be used to change the gender norms associated with such tasks. We design three treatments, where a) a volunteer receives positive social recognition, b) a non-volunteer receives negative social recognition, and c) a volunteer receives positive, but a non-volunteer receives negative social recognition. Our results indicate that competition for social recognition increases the overall likelihood that someone in a group has volunteered. Positive social recognition closes the gender gap observed in the baseline treatment, so does the combination of positive and negative social recognition. Our results, consistent with the prior literature on gender differences in competition, suggest that public recognition of volunteering can change the default gender norms in organizations and increase efficiency at the same time.
\end{abstract}


\end{titlepage}

%\end{}
\begin{doublespacing}

\clearpage

\newpage
\setcounter{page}{1}
\input{1Introduction}
\input{3Design}

%\input{3Predictions}
\input{4Results}

\input{5Conclusion}

\newpage
    \addcontentsline{toc}{section}{References}
    \bibliographystyle{aer}
    \bibliography{References.bib}
    %\bibliography{zotero.bib}

\input{6Appendix}

\end{doublespacing}
\newpage
\end{document}

%% file: 1Introduction.tex
\section{Introduction}\label{Section-Introduction}

The labor market outcomes remains systematically different between men and women, despite half a century's research-based policies \citep{altonji1999race, marianne2011new}. While some scholars have focused on the role of discrimination and differences in productivity, others have taken preference or belief -based approaches as key mechanisms driving the stubborn gender gap in such outcomes. These mechanisms can result in equilibria where the nature of tasks men and women end up performing are different. Such a separating equilibrium may be perpetuated for several reasons: managers may assign women fewer challenging tasks \citep{de2010gender} or women themselves may choose challenging tasks less often \citep{niederle2008gender}. 
\par 
Relatedly, in an influential recent study, \cite{babcock2017gender} use field data from a large public university to show that women are significantly more likely to volunteer in senate committees than men. Their findings are consistent with those of \cite{porter2007closer} who show that female faculty spend 15 percent more time on committees than their male counterparts. Such service-oriented tasks, such as volunteering in committees, are called low-promotability tasks because they matter less either for performance evaluation or for raising outside options. \cite{babcock2017gender} (BRVW, henceforth) complement these findings with laboratory experiments. These experiments mimic the incentive structure of low-promotability tasks where the person volunteering is put at a relative disadvantage, but others enjoy the benefits of the completed task. The paper shows that women indeed volunteer more than men in such tasks.\footnote{Examples of low-promotability tasks are organizing recreational events at offices or serving in committees in academic settings.} They trace the origin of this gender gap to differences in beliefs about volunteering and not preferences.
\par 
There are many potential negative consequences of the gender-gap in volunteering for low-promotability tasks. In the long run, this gap may result in systematic differences in outcomes across gender, such as a under-representation of women in leadership positions, a widening wage gap between men and women, and a differential selection out from competitive environments. Given these findings, a central question facing organizations is: how should the incentive structures be designed such that the gender-gap in volunteering for low-promotability tasks can be minimized?
\par 
The standard approach in economics to change behavior is to alter the economic incentive or, more specifically, monetary incentive. Notice that monetary incentives cannot be changed for low-promotability tasks since that will change the very definition of the task, i.e., the task will not remain a low-promotability one anymore\footnote{This is not to suggest that low-promotability tasks should not be rewarded by organizations, but to highlight that certain tasks will always rank lower in the hierarchy relative to others, in an organizational setting.}. Moreover, there is little evidence that there is a difference in the perceived value of monetary incentives across gender. If so, economic incentives can alter the level of volunteering for both genders without addressing the gender gap. In this paper, we investigate if competition for non-monetary incentives, such as positive and negative social recognition, can help close the gender gap in volunteering for low-promotability tasks.
\par 
We adapt the design of BRVW and conduct a between-subject laboratory experiment to answer the above question. In the baseline treatment, participants are randomly divided into groups of three and play ten rounds of an investment game -  a modified version of the volunteer's dilemma game. In each round, they decide, within a specific timeout, whether to invest (i.e. volunteer) or not by clicking on a button. If no one invests, then the group-members get the default payoff. If someone invests within the timeout, she earns a little more than the default payoff, but others earn more than her. Thus, the incentive structure in the game mimics real life situations where everyone wants others to perform a particular task. A design feature of the baseline treatment that distinguishes our study from BRVW is the following: participants begin the game by first choosing gender salient, fictitious names. The fictitious names of the investors are shown on the screens of the group members. In the positive social recognition treatment, investors' real name is flashed on the screens of the group members alongside a congratulatory message and a smiling emoji (\smiley{}). Besides, investors are publicly applauded through a pre-recorded audio. Since a name typically signals the gender of the person, socially recognizing a participant reveals the gender of the investor. This may have unintended spillover effects on investing from one round to the other. This is precisely why the investors in the baseline treatment are identified with a fictitious, gender salient name. The difference between the baseline and the social recognition treatments determines the uncontaminated effect of social recognition. 
\par 
In the negative social recognition treatment, the real names of non-investors are displayed on the screens of the group members with a frowning emoji (\frownie{}).  In the positive and negative social recognition treatment (\smiley{}\frownie{}), the real name of the investor is displayed on the screens of the group members with a congratulatory message and a smiley; in contrast, those of non-investors are displayed alongside a frownie. A comparison of the investment decision in the treatment conditions with the baseline helps identify the effect of social recognition on volunteering, which in turn allows us to comment on the gender gap in volunteering across different treatment conditions. Note that positive (negative) social recognition is a non-monetary incentive, which increases the net benefit (cost) of volunteering (not volunteering). To that extent, it is plausible that the social recognition interventions increase the overall chances of investment in a group, as any other standard incentive. Our experimental design allows us to identify that.   
\par
What are our main findings? In the baseline treatment, women invest significantly more than men, and the difference is significant at the 5\% level. Our baseline treatment replicates the gender-gap in volunteering for low-promotability tasks. Strikingly, the gender-gap is not significantly different from zero in the positive social recognition treatment and the treatment with both positive and negative social recognition. However, in the negative social recognition treatment, the gender gap is persistent. A period-wise representation of the investment decisions confirms that the gender-gap is persistently high over time in the baseline and the negative social recognition treatments, but not so in the other two treatments. 
While two of our interventions successfully eliminate the gender gap in volunteering, a key question remains. Are the interventions efficient, as measured by the proportion of groups where investment occurs?  Interestingly, the answer is yes - the proportion of groups where someone invests is significantly higher in the social recognition treatments relative to the baseline. Thus, not only does the \smiley{} and \smiley{}\frownie{} treatments eliminate the gender gap, they do so in a manner that increases overall efficiency.
\par
 Our paper makes several key contributions to the literature. First, to the best of our knowledge, ours is one of the first papers to successfully test a policy intervention that can close the gap in low-promotability tasks' performance. In a related paper, \cite{bacineeckel} show that it is possible to break the gender norm by introducing asymmetric public costs for volunteering. The asymmetric costs serve as the focal point around which volunteering is coordinated, and through that, close the gender gap. In contrast, this paper exploits gender differences in competition for non-monetary incentives. 
 \par 
 Second, we contribute to the literature on social recognition by showing that it is a useful policy tool through which entrenched gender norms can be altered. Field-based experiments have demonstrated that social recognition (or more generally, social incentive) improves performance \citep{ashraf2014no}, increases voter turnout \citep{gerber2008social}, increases blood donation \citep{lacetera2009will} and increases volunteering activities and employees' motivation \citep{gallus2017fostering, gallus2016awards}. Social comparison, on the contrary, reduces performance \citep{ashraf2014awards}.\footnote{For a review of the literature on social incentive, see \cite{ashraf2018social}.}  On the lab front, studies have shown that socially recognizing contributors increases the contribution levels in public good games \citep{andreoni2004public, rege2004impact}, social approval for energy conservation decreases energy consumption \citep{schultz2007constructive} and social visibility and self image concerns increase volunteering \citep{jones2014wallflowers,exley2018incentives}. While the literature primarily focuses on positive social recognition, \cite{samek2014recognizing} have shown that recognizing the non-contributors have the highest effect on contributions in public good games. We contribute to this literature by showing that in social dilemma environments, social recognition can play a significant role in helping society move towards more efficient outcome. 
\par 
Third, some recent studies have shown that the effect of social recognition may differ across gender.  \cite{jones2014wallflowers} demonstrate that women are more likely to exhibit the 'wallflower effect' - a tendency to respond under visibility in a manner which reduces scrutiny. Some find women have a greater sense of shame than men \citep{ludwig2017women}, a lower self-perception \citep{exley2019gender} and self-stereotype is one mechanism through which such gender differences perpetuate \citep{gallus2020awards}. Our research contributes to this literature by showing that women respond differently to social recognition than men. In fact, the root of this difference can be traced back to the literature on gender differences in competition \citep{niederlevester2007}, with the important difference being, men, in our case, compete more than women for a non-monetary reward - positive social recognition.  In other words, we exploit gender difference in one aspect of behavior to close the gender-gap in another. The precise mechanisms behind the gender difference in responses to social recognition are important open question for future research.
\par
The rest of the paper is organized as follows: Section 2 lays out the experimental design and procedure, Section 3 presents the results, while Section 4 offers the concluding remark.

%% file: 3Design.tex
\section{Experimental Design and Procedure}\label{Section-Design}
% Note: more ifnormation about the gender composition of the group in the neg social recog treatment
%describe the equilibrium in short
% what was the statement used in the negative social recog treatment.

We conduct a laboratory experiment in India to answer the research questions discussed above. The structure of the game captures the essence of the incentives faced by a small group that is looking for a volunteer for a task that everyone prefers undertaken, but none wants to undertake it. In other words, if no one volunteers the dominant strategy of a player is to volunteer. The baseline treatment is designed to identify the baseline gender gap in volunteering for such low-promotability tasks in our sample. The subsequent three treatments are designed to examine whether non-monetary incentives such as social recognition can help close this gender gap.

\subsubsection*{Baseline Treatment}

In the baseline treatment, individuals are randomly and anonymously assigned to groups of three, in each of the ten rounds of play.\footnote{The group assignment is done in a way such that the exact three members are not repeated twice in a row.} A group has T seconds within which a member can make the investment (or volunteer), where T is drawn randomly from a uniform distribution with support from 45 and 90. If no one invests in a particular group within the timeout, each member earns 100 experimental currency units or \textbf{M}ohars.\footnote{Mohars or gold coins were a medium of exchange in ancient India.} The moment one group member invests, the round ends. The investor receives a payment of \(300M\), while the other two group members each receive \(900M\). In case multiple participants invest simultaneously, the investor is determined randomly. When no one invests within the timeout, the group members are informed about it and are given $100 M$. The ordinal ranking of the payoffs are the same as in BRVW. Assuming, rational agents, the equilibria of the game comprises of a pure strategy Nash equilibrium where one player invests and others do not, a mixed strategy Nash equilibrium where the likelihood of investment is 0.232, a mixed strategy Nash equilibrium where one person does not invest and the other two invest with probability 0.40. Overall, the probability that someone in a group invests is 1, 0.54 or 0.64, depending on which equilibrium is played.
\par 
There are two key differences in our design of the baseline treatment with respect to BRVW. First, in their design, a group had 2 minutes to invest, but in our setup, the timeout happens randomly between 45 and 90 seconds.\footnote{The difference was motivated by the results from the pilot data, which suggested that investment decisions were bunched with little or no variation when the duration of the round was fixed. Also, given the bunching at the end, the absence of mouse-click meant participants knew others in the room had not invested-our design avoids such potential peer effect. Notice, with the uncertainty related to timeout, risk preference becomes a crucial explanatory variable, which we control for in our results.} Second, participants were described in neutral terms such as Participant A in BRVW and gender salience of the participants came from the fact that participants in the lab were only from one gender or mixed gender. In order to make sure that the gender information revealed is constant across treatments, we use fictitious, gendered names for participants in the baseline treatment. Before starting the experiment, participants are first asked about their gender (and age). Subsequently, they choose a name for themselves from a list of gender revealing names, commonplace in India, which they retain through the course of the session. The chosen names are accepted only if they are consistent with the reported gender. Once a name is matched to a participant, it is removed from the choice set and others cannot pick it anymore.\footnote{In a separate incentivized survey with 91 respondents, we elicit the perceptions about the gender associated with the names we use. The empirical likelihood of correct perception turned out to be 97.9\%.} Experiments where participants choose nicknames have been conducted earlier as well, sometimes with an aim to make the gender identity salient \citep{duersch2009incentives,wiborg2020collaboration}. Examples of names we used are Gautam and Rohan for males and Ananya and Sonia for females.\footnote{Given these were common Indian names,  what if there actually were session participants with these names? Would they feel socially recognized even in the baseline treatment? If they did, then the observed gender-gap in the baseline might be interpreted as a lower bound of the true gap.} If a group member invests in a round, then his/her fictitious name with the following message is shown on all three participants' screens: ``$<fictitious name>$ of your group invested in Round $<x>$". Besides this message, their own investment decision and the corresponding earnings in the round are displayed. 
\par 
This aspect of the design was necessitated by the following: the social recognition treatments described below reveal the real names of those who invest. However, names reveal not only the identity but also the gender. Consequently, the investment decisions in subsequent rounds may be colored by not only the effects of social recognition but also the gender of the past investors.{\footnote{Although one round is randomly picked for paying subjects, there are some recent evidence of behavioral spillovers because of information revealed in one round to behavior in another \citep{Banerjee2018567,Leibbrandt2018}.}} An alternative baseline treatment could have been one where the investors revealed their true names but were not socially recognized. However, it is hard to imagine that revealing the true name of the investor will not generate some form of social recognition. For these reasons, we chose a baseline treatment with fictitious names such that the gender of the investor is revealed, and the treatment is strategically identical to BRVW, but the investor is not socially recognized. Overall, this helps identify the pure effects of social recognition when compared with the other treatments.
\subsubsection*{Treatment \LARGE\smiley{}}
In the positive social recognition treatment, participants first respond to the survey, type their real names and then start playing the game. At the end of every round, the name of the investor is displayed on the screens of each group member along with a congratulatory message and a smiling emoji - \smiley{}. Along with this, an audio clip of applause is played, and the participants are encouraged to join the applause. Figure \ref{fig:psr} presents the screen where the investor is positively recognized.
\par
\subsubsection*{Treatment \LARGE\frownie{}}
 In the negative social recognition treatment, participants type their real names before they start playing the game. At the end of every round, the names of the two individuals who chose not to invest is displayed on the screens of each group member along with a frowning emoji - \frownie{}. Note, information about the investor is not communicated to the group members in this treatment. Figure \ref{fig:nsr} presents the screen where the non-investor receives negative recognition.

\subsubsection*{Treatment \LARGE\smiley{}\LARGE\frownie{}}
 In many practical situations, organizations may find it hard to implement a policy where non-volunteers are 'called out'. It may be much easier for organizations to publicize the names of both volunteers and non-volunteers. To test the effectiveness of such a policy, we conduct a final treatment where we display the names of the investor and the non-investors. At the end of each round, the investor's name is displayed with a congratulatory message and a \smiley{}, while the names of the two non-investors are displayed along with a \frownie{}. Figure \ref{fig:pnsr} presents the screen where the investor is positively recognized and the non-investors are negatively recognized.  \\

Thus, the four treatments we implement differ in the way the investor and/or the two non-investors are recognized within each group, after each round. Notably, the interventions discussed above change only the non-monetary payoff associated with volunteering, while leaving the strategic aspects of the game unchanged.

\subsubsection*{Experimental Procedure}

The experiment was coded up in z-Tree \citep{fischbacher2007z} and was conducted at a large private university in India. Participants were recruited from undergraduate classes and randomized into sessions. None of the participants had prior experience in economic experiments. Twenty-three sessions were conducted with participants between 9 and 24 in each. A total of 285 students participated (75 in Baseline, 75 in Treatment \smiley{}, 60 in Treatment \frownie{} and 75 in Treatment \smiley{}\frownie{}). Sessions were roughly gender-balanced with the share of women in a session ranging between 33 and 60 percent. The timer in each round was implemented only through an audio clip that counted seconds, which in turn helped mask the sound of mouse clicks. On screen timer was not used. At the end of the ten rounds, participants answered a short survey, which included questions on gender, age, religion, caste, average monthly family income and year in college. The participants earned an average of Rs. 170 ($\approx$10 PPP-USD) for a session which lasted about one and a half hours.

%\footnote{The population was homogeneous. The average demographics and whether they differed significantly from the Baseline treatment is reported in Table \ref{tab:summary_stats}} 

%% file: 4Results.tex
\section{Results}\label{Section-Results}

%%Summary stats
Table \ref{tab:summary} reports the descriptive statistics of the outcomes along with the behavioral and demographic control variables and compares them across treatments. The behavioral control variables, namely, altruism, risk, non-conformity, and agreeableness are balanced across treatments. A \textit{t}-test with respect to the baseline reveals only Altruism is significantly different in Treatment \frownie{} from the baseline at 10\% level. The demographic control variables are generally well balanced, too, with only Age in Treatment \smiley{} significantly different from the baseline, at 10\% level and Years in College in Treatment \smiley{} and \frownie{} are significantly different from the baseline at 5\% level. In the regression analysis, we control for these variables.

\begin{figure}[H]
   \caption{Comparing Individual and Group Investment across Treatments}
     \centering
     \begin{subfigure}[b]{0.45\textwidth}
         \centering
         \includegraphics[width=\textwidth]{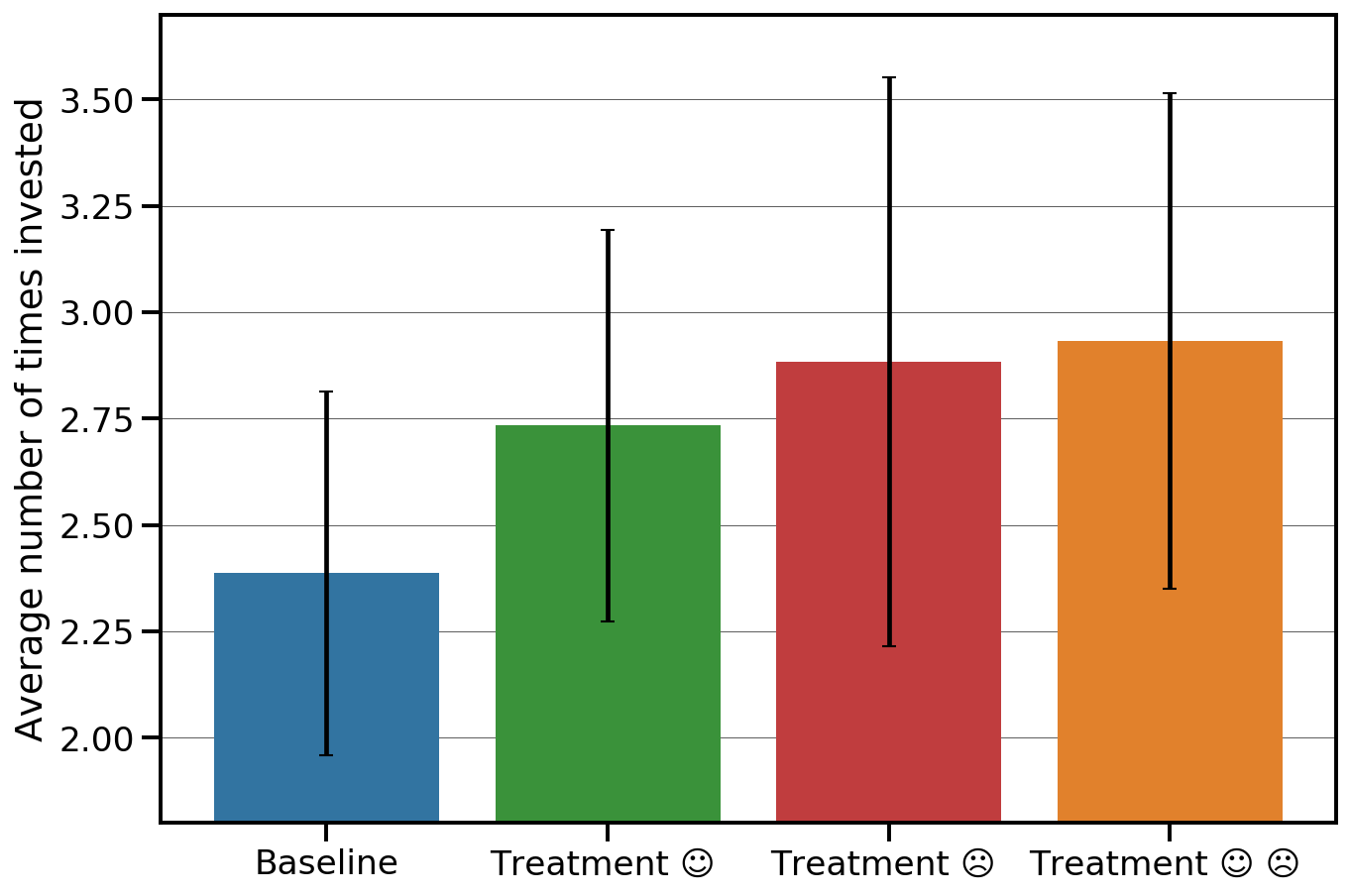}
         \subcaption{Mean \(TotalInvestment\) across treatment}
\footnotesize The average Investment is 2.39 in the Baseline, while it is 2.73 (\(t\)-test, \(p\)-value=0.27) in Treatment \smiley{}, in Treatment \frownie{} it is 2.88 (\(t\)-test, \(p\)-value=0.20) and  2.93 (\(t\)-test, \(p\)-value=0.13) in Treatment \smiley{}\frownie{}.
     \end{subfigure}
     \hfill
     \begin{subfigure}[b]{0.45\textwidth}
         \centering
         \includegraphics[width=\textwidth]{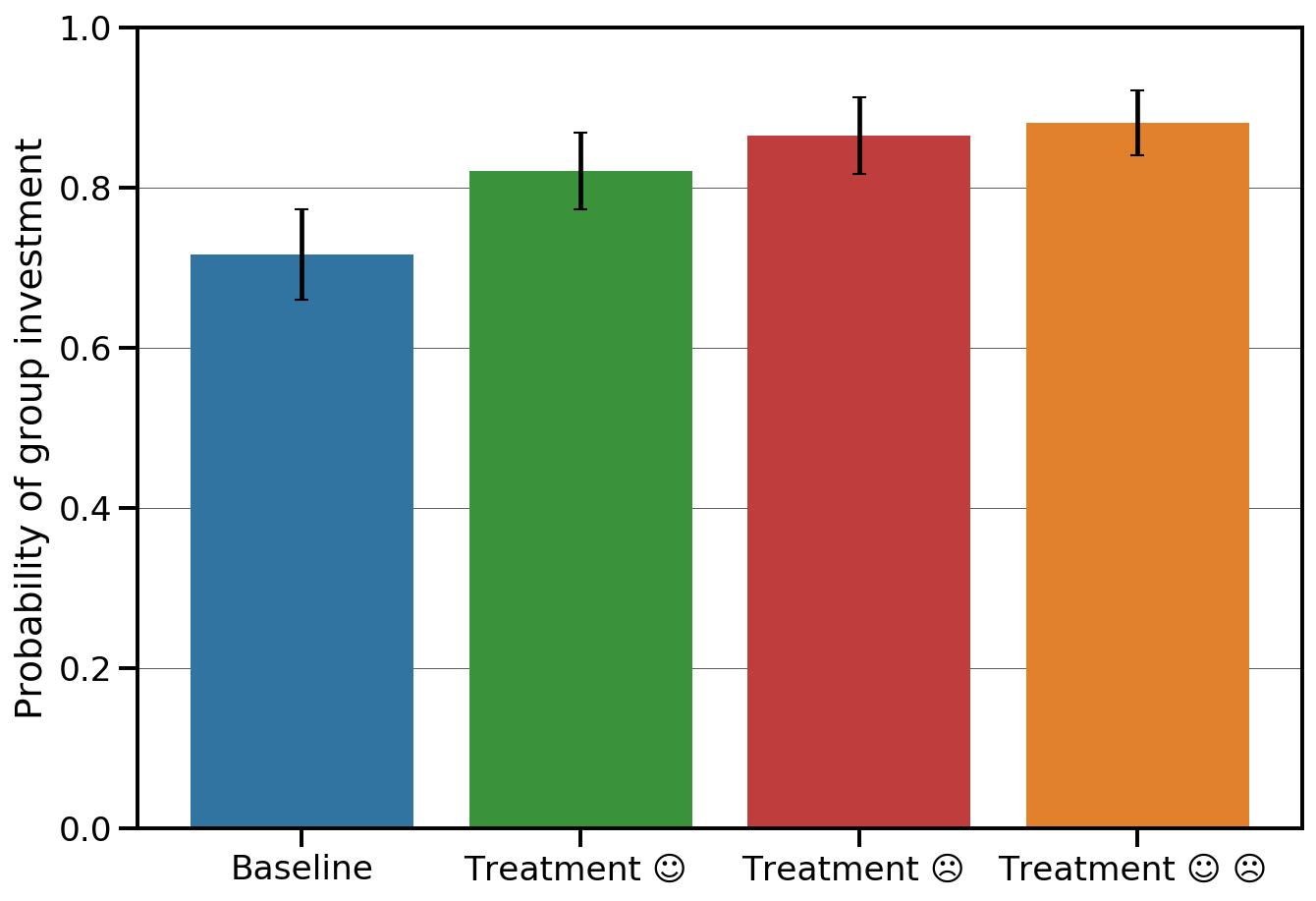}
         \subcaption{Mean GroupInvestment across treatment}
\footnotesize The average GroupInvestment is 71.6\% in the Baseline while it is 83\% in Treatment \smiley{} (Fisher exact test, \(p\)-value$<$0.01), 86\% (Fisher exact test, \(p\)-value$<$0.01) and 87\% for Treatment \frownie{} (Fisher exact test, \(p\)$<$0.01).
     \end{subfigure}
\label{fig:indgroupinv} 
\end{figure}

We start the analysis by comparing the average investment per participant across the treatments. Figure \ref{fig:indgroupinv}(a) reports the average number of times an individual invested in the ten rounds. The social recognition treatments increase investment, but the difference is not statistically significant when compared to the baseline. Figure \ref{fig:indgroupinv}(b) plots the average number of times an investment takes place in a group, across the treatments. The group investment increases from 71.6\% in the baseline to 82\% in the \smiley{} treatment (Fisher's exact test, \textit{p}-value$<$0.01), 86\% in the \frownie{} treatment (Fisher's exact test, \textit{p}-value$<$0.01) and 88\% in the \smiley{}\frownie{} treatment (Fisher's exact test, \textit{p}$<$0.01). These results hold in a regression framework which controls for demographic and behavioral variables and clusters standard errors at the session level, as reported in Table \ref{tab:indgroupinv}. The \smiley{} treatment dummy is positive and significant at 5\% level, while \frownie{} and \smiley{}\frownie{} treatment dummies are significant at 1\% level.  Overall, the group investments are consistently higher in the social recognition treatments relative to the baseline. This indicates that socially recognizing voluntary contributions in organizational settings can play the role of an effective non-monetary incentive. The period-wise probability of individual investment and group investment are plotted in Figure \ref{fig:ind_group_inv_distribution}(a) and \ref{fig:ind_group_inv_distribution}(b), respectively. Interestingly, the proportion of groups in the baseline treatment where investment occurs is significantly greater than either 0.54 or 0.64 and lower than 1 - the equilibrium predictions (\(t\)-test, \(p\)-values $<$0.01,
$<$0.01, $<$0.01, respectively). Thus, these is some suggestive evidence that participants in our sample are either behavioral or hold different views about which of the equilibria will be played by the other participants. We now have our first set of results:
\begin{result}
Social recognition does not significantly increase the rate at which an individual volunteers. 
\end{result} 

\begin{result}
Social recognition significantly increases the likelihood that someone in the group volunteers. 
\end{result} 

\begin{comment}
\begin{figure*}
  \centering
  \begin{tabular}{c @{\qquad} c }
    \includegraphics[width=.45\linewidth]{Figure1a.png} &
    \includegraphics[width=.45\linewidth]{Figure1b.png}\\
    \small (a) Mean TotalInvestment across treatment & \small (b) Mean GroupInvestment across treatment
  \end{tabular}
  \caption{Comparing Individual and Group Investment across Treatments}
\end{figure*}
\end{comment}

% Treatment differences in the likelihood of individual investment and group investment
\begin{table}[hbt!]
  \centering
   \caption{Treatment Differences in the Likelihood of Individual Investment and Group Investment \label{tab:indgroupinv}}
  \begin{threeparttable}
    \begin{tabular}{lccccc}
    %\multicolumn{7}{c}{Table 1: Probability of Individual Investment and Group Investment} \\
    \toprule
   % \toprule
     & \multicolumn{3}{c}{Investment}  & \multicolumn{2}{c}{GroupInvestment} \\
    \cmidrule(lr){2-4} \cmidrule(lr){5-6}  
    \multicolumn{1}{l}{} & (1)   & (2)   & (3)   & (4)   & (5)    \\

\midrule

   Treatment \Large\smiley{} & 0.0365 & 0.0330 & 0.0291 & 0.0822** & 0.0706** \\
          & (0.0334) & (0.0339) & (0.0347) & (0.0407) & (0.0342) \\
    Treatment \Large\frownie{} & 0.0520 & 0.0521 & 0.0543 & 0.118*** & 0.108*** \\
          & (0.0418) & (0.0411) & (0.0412) & (0.0314) & (0.0213) \\
    Treatment \Large\smiley{}\Large\frownie{} & 0.0568 & 0.0590 & 0.0547 & 0.135*** & 0.125*** \\
          & (0.0380) & (0.0369) & (0.0377) & (0.0293) & (0.0220) \\
    Round & -0.00355 & -0.00355 & -0.00351 &       & -0.0106** \\
          & (0.00258) & (0.00260) & (0.00260) &       & (0.00419) \\
    Female Session Share &       &       & 0.0410 &       & 0.345*** \\
          &       &       & (0.131) &       & (0.118) \\
          &       &       &       &       &  \\
    Observations & 2,850 & 2,850 & 2,850 & 950   & 950 \\
    Behavioral Controls & No    & Yes   & Yes   & -      & -  \\
    Demographic Controls & No    & No    & Yes   & -      & -  \\
    Cluster & Individual & Individual & Individual & Session & Session \\
    \bottomrule
    \end{tabular}%

      \begin{tablenotes}
            \item[\textdagger] \footnotesize The table presents marginal effects. Standard errors are reported in parentheses, clustered as individual level for col (1)-(3) and session level for col (4)-(5). Treatment dummies = 0 for baseline. The dependent variable for Col(1)-(3) is Investment (1-invest, 0-don’t invest). Behavioral controls include controls for preferences for risk , altruism, non-conformity and agreeableness. Demographic controls include controls for age, religion, caste, years in college and family income categories. With group as the unit of observation, the dependent variable for col (4)-(5) is GroupInvestment (1-atleast one investor in group, 0-no investor in group). 
            *** p$<$0.01, ** p$<$0.05, * p$<$0.1
        \end{tablenotes}
   \end{threeparttable}

\end{table}

We now come to the main result in our paper. Figure \ref{fig:gendergap} plots the probability of investment by gender across treatments. In the following analysis, we statistically compare the count variable, \(TotalInvestment\), across treatment, and gender. In the baseline treatment, on average, males invest 1.9 times, while females invest 2.8 times. The gender gap in the number of times participants invest is 0.9 and is statistically significant (\textit{t}-test, \textit{p}-value=0.038). The median \(TotalInvestment \) for males and females are 2 and 2.5, respectively (Mann Whitney test, \(p\)-value=0.045).  Table \ref{tab:gendergap} regresses the dummy variable \textit{Investment} on \textit{Female} using a probit model. The most stringent version of the model includes behavioral and demographic controls, session dummies and clusters standard error at the individual level. The behavioral controls, constructed in the same way as in BVWR, are: risk, altruism, non-conformity and agreeableness. In addition, we control for the share of females in a session to account for the behavioral responses owing to the variation in the number of women in a session. The demographic controls are: age, religion, caste, years in college and family income categories.\footnote{For variable definitions and treatment averages, refer to Table \ref{tab:summary}.} Through this, we match the exact specification reported in BVWR.  The coefficient corresponding to \textit{Female} is positive and significant, suggesting that the females volunteer significantly more than men do. This gender gap is also clear when one looks at the probability of investing over period in Figure \ref{fig:gendergap_time}(a).\footnote{Additional robustness checks with timeout as a control variable does not change the result (not reported).} Thus, we closely follow BRVW in terms of our experimental design and empirical specification and replicate the gender gap in the likelihood of volunteering.
\begin{result}
Women volunteer significantly more than men in the absence of any form of social recognition. 
\end{result} 
\setcounter{figure}{1}

\begin{figure}[htbp]   
\centering

\includegraphics[scale = 0.4]{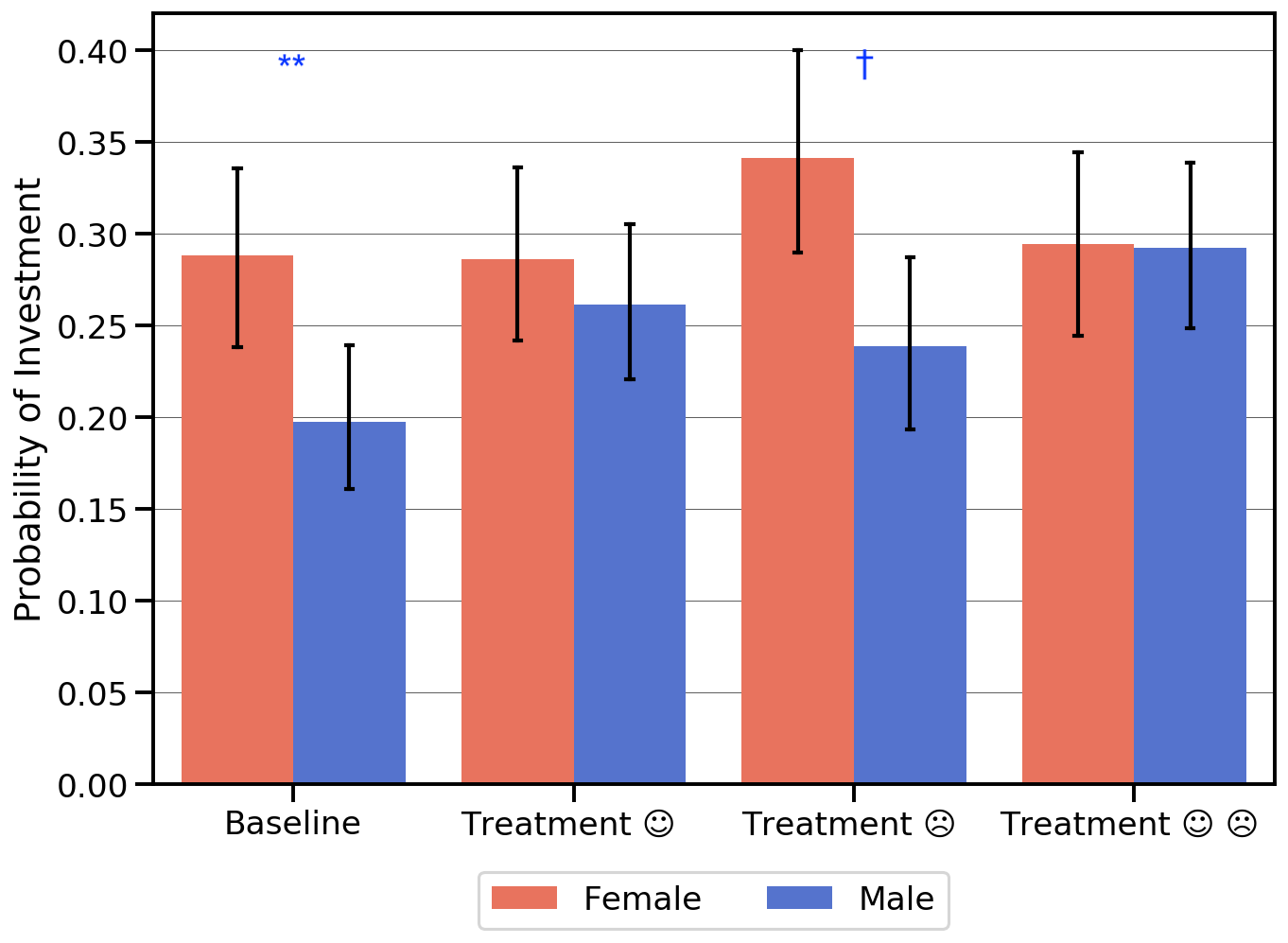}    
\caption{Gender gap in investment across treatments} 
\label{fig:gendergap}
\floatfoot{The figure plot the probability of \(Investment\), separately for each gender across the treatments. In the baseline treatment, females invest 2.88 times, while males invest 1.98 times (\(t\)-test, \(p\)-value=0.03). In Treatment \smiley{} females invest 2.86 times while males invest 2.62 times (\(t\)-test, \(p\)=0.60). In Treatment \frownie{} females invest 3.41 times while males invest 2.39 times (\(t\)-test, \(p\)-value=0.13). In treatment \smiley{}\frownie{} females and males invest 2.92 and 2.94, respectively (\(t\)-test, \(p\)-value=0.97). The equality of probability of \(Investment\) between male and female is compared for each treatment. *** p$<$0.01, ** p$<$0.05, * p$<$0.1,\dagger p$=$0.13.}
\end{figure}
\par 
In the \smiley{} treatment, the picture is quite different. Females invest 2.86 times while males invest 2.62 times (\textit{t}-test, \textit{p}-value=0.60). The median \(TotalInvestment \) for males and females are 2 and 3, respectively (Mann Whitney test, \(p\)-value=0.51). The corresponding regression results are reported in Columns (4)-(6) of Table \ref{tab:gendergap}, where the most stringent specification, model (6), includes behavioral and demographic controls and session dummies. The coefficient of females is consistently statistically insignificant. The fact that the gender gap is closed is also clear from Figure \ref{fig:gendergap_time}(b). This gives us our next result.
\begin{result}
In the presence of positive social recognition for volunteers, there is no difference in the likelihood of volunteering between men and women.  
\end{result} 
Thus, when volunteers receive positive social recognition, the erstwhile gender gap disappears. In the \frownie{} treatment, females invest 3.41 times on an average, while males invest only 2.39 times. While the raw gender gap is not statistically significant at the 10\% level (\textit{t}-test, \textit{p}-value=0.13), column (9) of Table \ref{tab:gendergap} presents the coefficient of \textit{Female} after controlling for behavioral and demographic variables and sessions. The coefficient of Female is positive and significant at 5\% level.\footnote{The raw gender-gap is larger in the \frownie{} treatment than in Baseline, so is the standard error. The latter makes the gender-gap statistically insignificant.} Further, the median \(TotalInvestment \) for males and females are 1 and 3, respectively (Mann Whitney test, \(p\)-value=0.03). The mean and median comparisons suggest that negative social recognition is unable to close the gender gap in volunteering, a finding which is supported by Figure \ref{fig:gendergap_time}(c). 
\begin{result}
Women volunteer significantly more than men in the presence of negative social recognition for non-volunteers. 
\end{result} 
Finally, in \smiley{}\frownie{} treatment, on average, the females invest 2.94 times while the males invest only 2.92 times. Once again, the gender gap is statistically insignificant (\textit{t}-test, \textit{p}-value=0.97), indicating that recognizing investors positively and non-volunteers negatively close the gender gap in our setup. The median \(TotalInvestment \) for males and females are 2 and 3, respectively (Mann Whitney test, \(p\)-value=0.91). The corresponding regression results reported in col (10) - (12) of Table \ref{tab:gendergap} show that the coefficient of Female is statistically insignificant. This is further confirmed by Figure \ref{fig:gendergap_time}(d).
\begin{result}
There is no difference in the likelihood of volunteering between men and women when volunteers are positively socially recognized, and non-volunteers are negatively socially recognized. 
\end{result} 
\par
Note that our experiment is not sufficiently powered to answer, for instance, whether males invest when positive social recognition is offered compared to the baseline or whether females invest more under negative social recognition than under positive social recognition. 
\par 
How do the treatments affect volunteering decisions among men and women? A cursory look at the data reveals that men invest more in \smiley{} than in Baseline, however, the difference is not statistically significant (\textit{t}-test, \textit{p}-value=0.13).  They do not invest significantly more in the \frownie{} than in Baseline, either. However, in the \smiley{}\frownie{} treatment, men invest significantly more than in Baseline (\(t\)-test, \(p\)-value=0.04). These results are consistent with the robust finding in the literature that, in general, men compete more than women \citep{niederlevester2007,niederle2011gender}. In this case, men compete more, not for monetary, but non-monetary rewards. Generally, competition for non-monetary rewards have been found to affect behavior such as altruism \citep{duffy2010does}. Ours is a case in the point. Women, on the other hand, invest more in the \frownie{} treatment than in Baseline, but the difference is not statistically significant (\(t\)-test, \(p\)-value=0.32). The difference in volunteering for women between the other treatments with respect to Baseline is minimal. Overall, the data offers some suggestive evidence that men react more to positive social recognition, while women tend to respond more to negative social recognition. The latter is consistent with earlier studies which find a greater sense of shame among women \citep{ludwig2017women}. Put together, the findings suggest that a combination of positive and negative social recognition may be an optimal strategy to increase volunteering in organizational settings. However, this is only suggestive and future research should aim to uncover the precise within-gender effects of social recognition. In addition, future research should also investigate the role of asymmetry in preference for social recognition among males and females and examine if such asymmetries can be used to plug gender-gaps in other areas.

\vspace{0.75cm}
\begin{figure}   

\includegraphics[scale = 0.4]{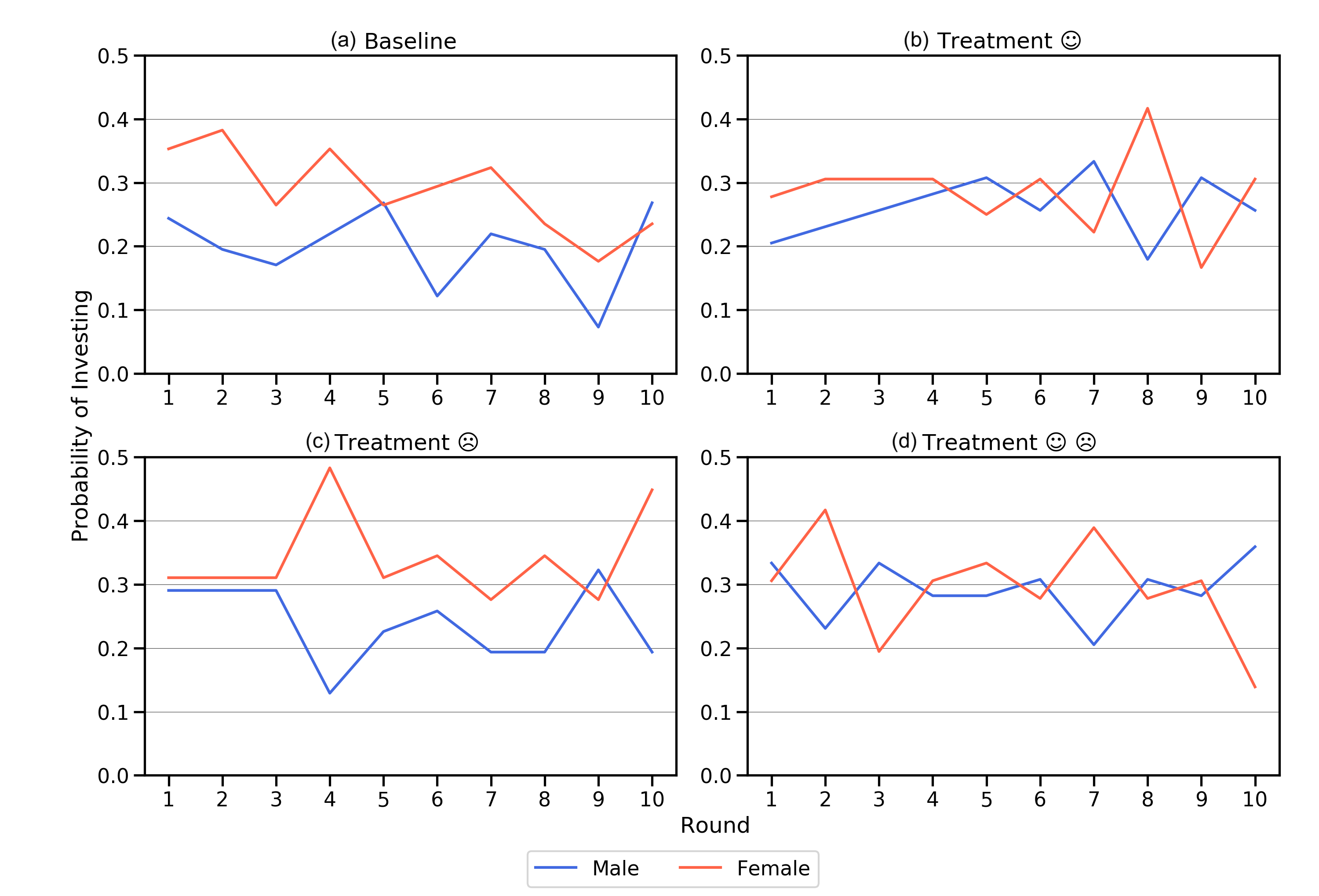}    
\caption{Gender gap in investment across treatment over round}    
\floatfoot{ The figure plots the gender-gap across treatment over period.}
\label{fig:gendergap_time}
\end{figure}

\begin{landscape}

% Gender Differences in Probability of Investment Across Treatment 
\begin{table}[h!]
     \caption{Gender Differences in Probability of Investment Across Treatment \label{tab:gendergap}}
\begin{threeparttable}
  \centering
  \scriptsize
    \begin{tabular}{lcccccccccccc}
    \toprule
          & (1)   & (2)   & (3)   & (4)   & (5)   & (6)   & (7)   & (8)   & (9)   & (10)  & (11)  & (12) \\
    \midrule
    \midrule
          &       &       &       &       &       &       &       &       &       &       &       &  \\
    Female & 0.0904** & 0.0866** & 0.0940* & 0.0246 & 0.0261 & 0.0190 & 0.103 & 0.0835 & 0.137** & 0.00197 & -0.0364 & -0.0307 \\
          & (0.0427) & (0.0429) & (0.0514) & (0.0460) & (0.0457) & (0.0495) & (0.0650) & (0.0624) & (0.0584) & (0.0587) & (0.0580) & (0.0583) \\
    Round & -0.00952* & -0.00959* & -0.00969* & 0.00124 & 0.00112 & 0.00113 & -0.00122 & -0.00137 & -0.00154 & -0.00404 & -0.00412 & -0.00416 \\
          & (0.00572) & (0.00569) & (0.00567) & (0.00526) & (0.00526) & (0.00526) & (0.00434) & (0.00449) & (0.00468) & (0.00490) & (0.00502) & (0.00510) \\
    Female Share Session &       &       & 0.00495 &       &       & 0.135 &       &       & -1.580 &       &       & -0.687 \\
          &       &       & (0.201) &       &       & (0.304) &       &       & (1.494) &       &       & (0.802) \\
          &       &       &       &       &       &       &       &       &       &       &       &  \\
    Observations & 750   & 750   & 750   & 750   & 750   & 750   & 600   & 600   & 600   & 750   & 750   & 750 \\
    Treatment & \multicolumn{1}{c}{Baseline} & \multicolumn{1}{c}{Baseline} & \multicolumn{1}{c}{Baseline} & \multicolumn{1}{c}{\Large\smiley{}} & \multicolumn{1}{c}{\Large\smiley{}} & \multicolumn{1}{c}{\Large\smiley{}} & \multicolumn{1}{c}{\Large\frownie{}} & \multicolumn{1}{c}{\Large\frownie} & \multicolumn{1}{c}{\Large\frownie} & \multicolumn{1}{c}{\Large\smiley{}\Large\frownie{}}& \multicolumn{1}{c}{\Large\smiley{}\Large\frownie{}}& \multicolumn{1}{c}{\Large\smiley{}\Large\frownie{}}\\
    Behavioral Controls & No    & Yes   & Yes   & No    & Yes   & Yes   & No    & Yes   & Yes   & No    & Yes   & Yes \\
    Demographic Controls & No    & No    & Yes   & No    & No    & Yes   & No    & No    & Yes   & No    & No    & Yes \\
    Session FE & No    & No    & Yes   & No    & No    & Yes   & No    & No    & Yes   & No    & No    & Yes \\
    \bottomrule
    \end{tabular}%
    \begin{tablenotes}
            \item[\textdagger] \scriptsize Dependent variable: individual investment decision (1-invest, 0-don’t invest). The table presents marginal effects. Standard errors are clustered at the individual level and reported in parenthesis. Behavioral controls include controls for preferences for risk , altruism, non-conformity and agreeableness. Demographic controls include controls for age, religion, caste, years in college and family income categories. *** p$<$0.01, ** p$<$0.05, * p$<$0.1
        \end{tablenotes}
 \end{threeparttable}
\end{table}%
\end{landscape}

%% file: 5Conclusion.tex
\section{Discussion and Conclusion}\label{Section-Conclusion}

\cite{babcock2017gender}'s influential work documents that women volunteer more often than men for tasks that are less likely to influence their evaluation and performance bonuses. Not only that, women also get asked to volunteer for low-promotability tasks more often. A key question is how institutional mechanisms can be put in place such that the gender gap in volunteering in low-promotability tasks can be reduced. 
\par 
We conduct a novel laboratory experiment to test whether socially recognizing volunteers for their effort can reduce the gender gap in volunteering. Besides, we also test the effectiveness of negative social recognition for non-volunteers and positive and negative social recognition for volunteers and non-volunteers, respectively. Our results indicate that positive social recognition and both positive and negative social recognition close the gender gap in volunteering; however, negative social recognition does not. Overall, the interventions prove efficient: increase the likelihood that at least someone volunteers for such tasks.
\par
  
\par
Our paper offers key policy recommendations to organizations. Any organization has tasks which directly add value to the objective it primarily cares about, and tasks which need to be done, but does not directly contribute to the organization's main objective. Consequently, the latter tasks are less rewarding and inevitably attract fewer people. Non-monetary incentives, such as social recognition, can increase the relative attractiveness of such tasks without changing the monetary payoff structure and in the process help close gender-gaps in volunteering for such tasks. Recent research indicates that there are significant welfare effects of such non-monetary policy interventions \citep{butera2019measuring}. Future research should investigate the welfare effects through through the lens of gender.

%% file: 6Appendix.tex
\renewcommand{\thesection}{\Alph{section}}
\setcounter{section}{0}
\setcounter{table}{0}
\setcounter{figure}{0}
%\captionsetup{figurewithin=section}
\renewcommand\thefigure{\Alph{section}\arabic{figure}}
%\captionsetup{tablewithin=section}
\renewcommand\thetable{\Alph{section}\arabic{table}}

\section{Appendix}\label{Section-Appendix}

\begin{landscape}

\begin{table}[h!]
   \caption{Descriptive Statistics} 
 \label{tab:summary}
  \begin{threeparttable}
  \centering
    \footnotesize
    \begin{tabular}{lrrrrr}
    \midrule
    \textbf{Variable} & \multicolumn{1}{l}{\textbf{Description}} & \multicolumn{1}{l}{\textbf{Baseline}} & \multicolumn{1}{l}{\textbf{Treatment \Large\smiley{}}} & \multicolumn{1}{l}{\textbf{Treatment \Large\frownie{}}}& \multicolumn{1}{l}{\textbf{Treatment \Large\smiley{}\Large\frownie{}}}\\
    \midrule
    \midrule
          &       &       &       &       &  \\
    \midrule
    \textbf{Outcome} &       &       &       &       &  \\
    \midrule
    TotalInvestment & \multicolumn{1}{p{19em}}{Total number of times an individual invested i.e. [0,10]} & \multicolumn{1}{c}{2.39} & \multicolumn{1}{c}{2.73} & \multicolumn{1}{c}{2.88} & \multicolumn{1}{c}{2.93} \\
    Investment & \multicolumn{1}{p{19em}}{=1 if individual invested in any particular round} & \multicolumn{1}{c}{0.239} & \multicolumn{1}{c}{0.273} & \multicolumn{1}{c}{0.288} & \multicolumn{1}{c}{0.293} \\
    GroupInvestment & \multicolumn{1}{p{19em}}{=1 if someone from the group invested in a round} & \multicolumn{1}{c}{0.72} & \multicolumn{1}{c}{0.82***} & \multicolumn{1}{c}{0.87***} & \multicolumn{1}{c}{0.88***} \\
    \midrule
    \textbf{Controls} &       &       &       &       &  \\
    \midrule
    Female Share Session & \multicolumn{1}{l}{Share of female subjects in any session} & \multicolumn{1}{c}{0.453} & \multicolumn{1}{c}{0.48} & \multicolumn{1}{c}{0.483} & \multicolumn{1}{c}{0.48} \\
          &       &       &       &       &  \\
    \midrule
    \textbf{Behavioral Controls} &       &       &       &       &  \\
    \midrule
    Altruism & \multicolumn{1}{l}{Index using question 9 to 11 in Survey} & \multicolumn{1}{c}{4.18} & \multicolumn{1}{c}{4.06} & \multicolumn{1}{c}{3.99*} & \multicolumn{1}{c}{4.11} \\
    Risk  & \multicolumn{1}{l}{Number of safe choices in Holt and Laury risk game} & \multicolumn{1}{c}{5.45} & \multicolumn{1}{c}{5.79} & \multicolumn{1}{c}{5.37} & \multicolumn{1}{c}{5.33} \\
    Non-Conformity & \multicolumn{1}{l}{Index using question 6 to 8 in Survey} & \multicolumn{1}{c}{3.61} & \multicolumn{1}{c}{3.62} & \multicolumn{1}{c}{3.8} & \multicolumn{1}{c}{3.56} \\
    Agreeableness & \multicolumn{1}{l}{Index using question 12 to 20 in Survey} & \multicolumn{1}{c}{3.76} & \multicolumn{1}{c}{3.68} & \multicolumn{1}{c}{3.79} & \multicolumn{1}{c}{3.75} \\
     %     &       &       &       &       &  \\
    \midrule
    \textbf{Demographic Controls} &       &       &       &       &  \\
    \midrule
    Age   & \multicolumn{1}{l}{in years} & \multicolumn{1}{c}{19.4} & \multicolumn{1}{c}{19.81*} & \multicolumn{1}{c}{19.52} & \multicolumn{1}{c}{19.49} \\
    Female & \multicolumn{1}{l}{=1 if subject is female} & \multicolumn{1}{c}{0.45} & \multicolumn{1}{c}{0.48} & \multicolumn{1}{c}{0.48} & \multicolumn{1}{c}{0.48} \\
    Caste & \multicolumn{1}{p{19em}}{=1 General, =2 SC , =3 ST , 4 OBC , 5 if Prefer not to say} & \multicolumn{1}{c}{1.47} & \multicolumn{1}{c}{1.44} & \multicolumn{1}{c}{1.52} & \multicolumn{1}{c}{1.84} \\
    Religion & \multicolumn{1}{p{19em}}{=1 Hindu , =2 Muslim , = 3 Christian , =4 Prefer not to say} & \multicolumn{1}{c}{1.57} & \multicolumn{1}{c}{1.71} & \multicolumn{1}{c}{1.58} & \multicolumn{1}{c}{1.47} \\
    Family Income & \multicolumn{1}{p{19em}}{=1: EWS , =2: LIG , =3: MIG , =4: HIG , =5: Rich , =6: Super Rich} & \multicolumn{1}{c}{4.05} & \multicolumn{1}{c}{4.31} & \multicolumn{1}{c}{3.93} & \multicolumn{1}{c}{3.83} \\
    Year in College & \multicolumn{1}{p{19em}}{=1 UG 1st , =2 UG 2nd , =3 UG 3rd , =4 UG 4th} & \multicolumn{1}{c}{2.36} & \multicolumn{1}{c}{2.52***} & \multicolumn{1}{c}{2.23**} & \multicolumn{1}{c}{2.43} \\
     %     &       &       &       &       &  \\
    \midrule
    \textbf{Sample Size} &       &       &       &       &  \\
    \midrule
    Sessions &       & \multicolumn{1}{c}{6} & \multicolumn{1}{c}{6} & \multicolumn{1}{c}{5} & \multicolumn{1}{c}{6} \\
    Subjects &       & \multicolumn{1}{c}{75} & \multicolumn{1}{c}{75} & \multicolumn{1}{c}{60} & \multicolumn{1}{c}{75} \\
    Rounds & \multicolumn{1}{l}{Number of times the game is played} & \multicolumn{1}{c}{10} & \multicolumn{1}{c}{10} & \multicolumn{1}{c}{10} & \multicolumn{1}{c}{10} \\
    Observations &       & \multicolumn{1}{c}{750} & \multicolumn{1}{c}{750} & \multicolumn{1}{c}{600} & \multicolumn{1}{c}{750} \\
    \midrule
    *** p$<$0.01, ** p$<$0.05, * p$<$0.1 &       &       &       &       &  \\
    \end{tabular}%
    \begin{tablenotes}
            \item[\textdagger] \footnotesize This table reports the mean of each of our control variables. A t-test is used to test statistical significance for difference between Treatment Baseline and each of the recognition treatments.
        \end{tablenotes}
   \end{threeparttable}
\end{table}
\end{landscape}

%Figures

%\begin{figure}
   % \centering
      %  \includegraphics[width=0.65\textwidth]{Figure A1.png} 
        %\caption{Distribution of Investment across Treatment}
 %\end{figure}      

\begin{figure}[h!]
    \centering
    \begin{subfigure}[b]{0.45\textwidth}
        \includegraphics[width=\textwidth]{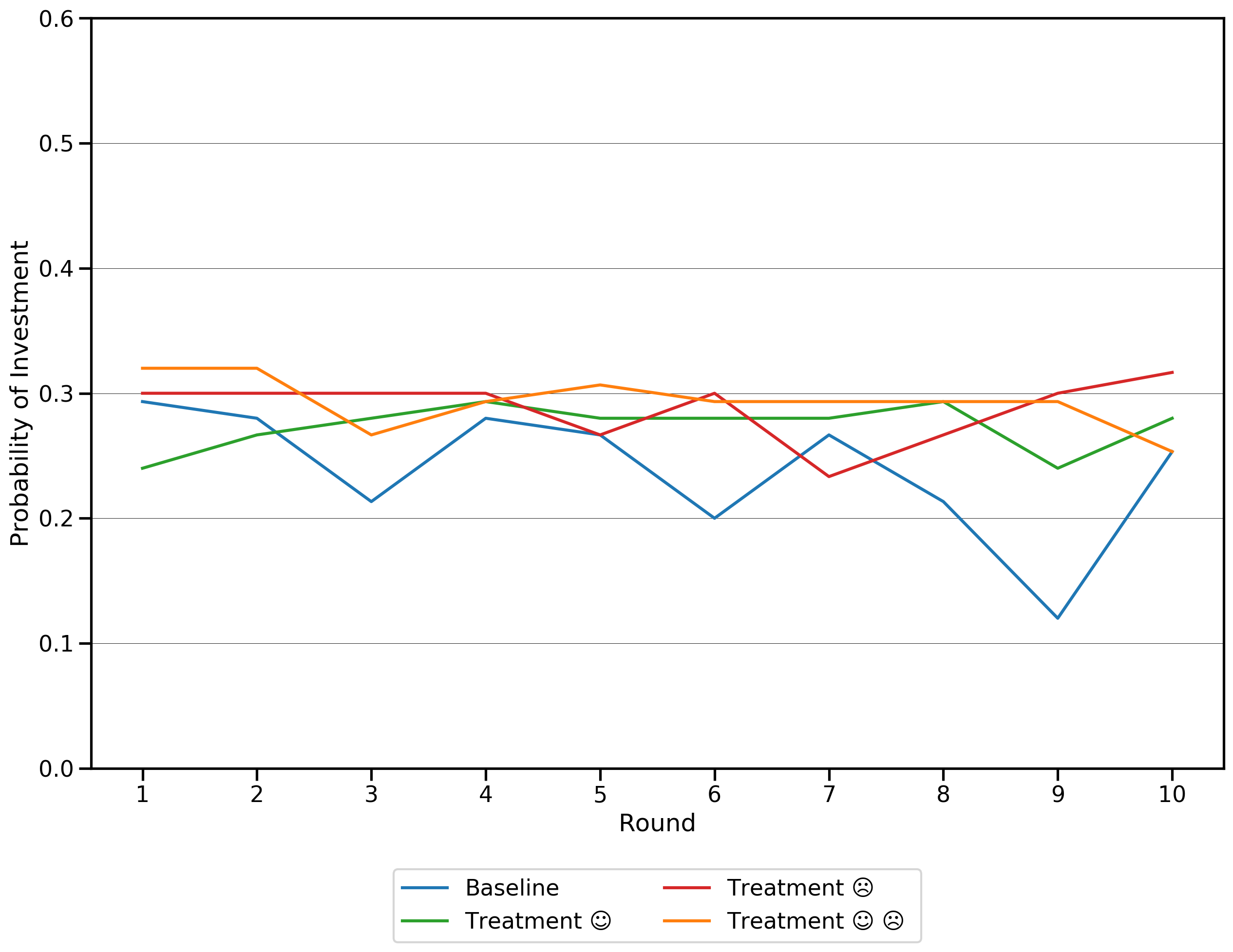} 
        \subcaption{Probability of Investment over round}
    \end{subfigure}
\quad
    \begin{subfigure}[b]{0.45\textwidth}
    \centering
        \includegraphics[width=\textwidth]{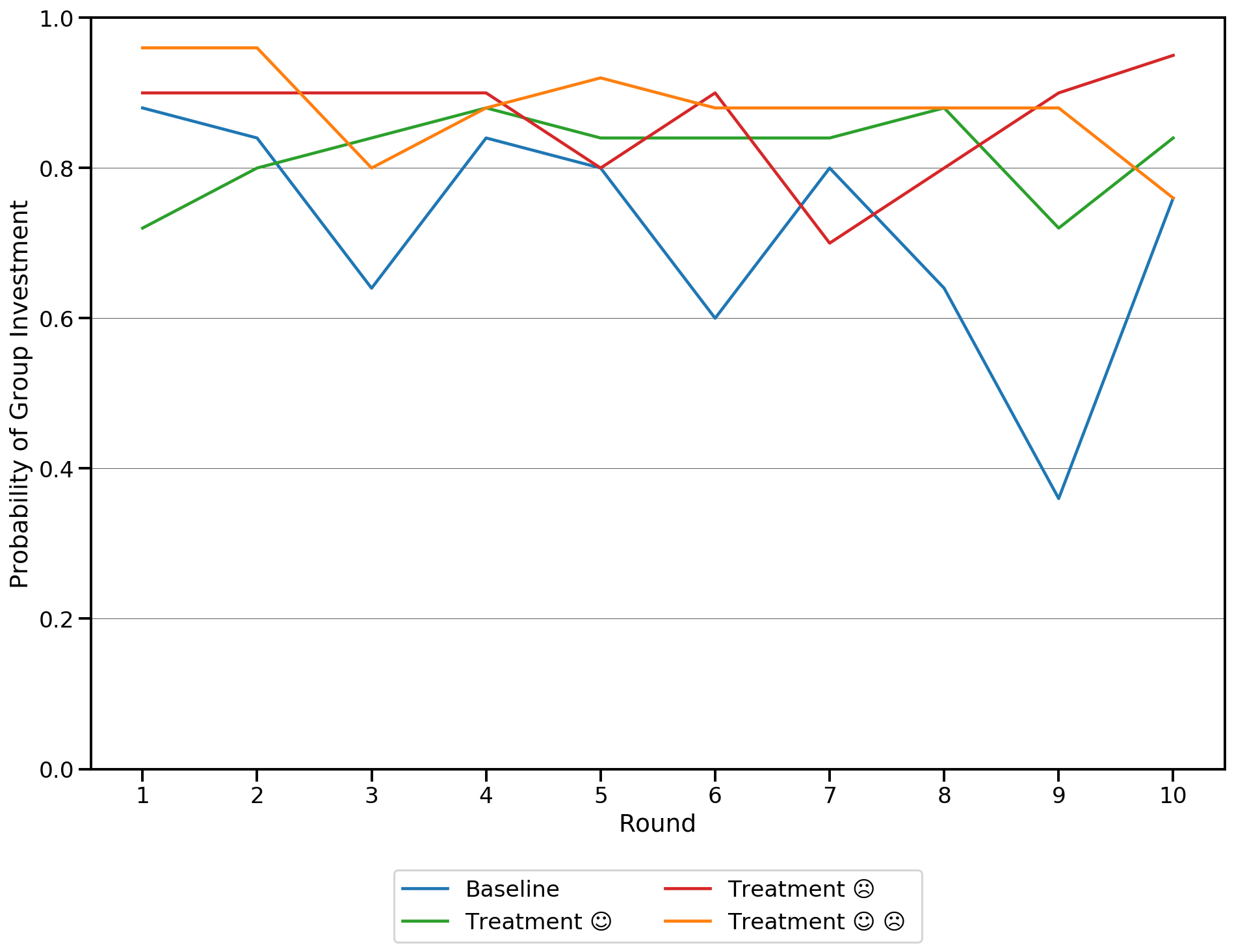}
        \subcaption{Probability of Group Investment over round}
        \label{fig:group_investment_over_period}
    \end{subfigure}
     \caption{Comparing Individual and Group Investment over round and across treatment}
      \label{fig:ind_group_inv_distribution}
\end{figure}    

\begin{figure}[h!]    
\centering
        \includegraphics[width=\textwidth]{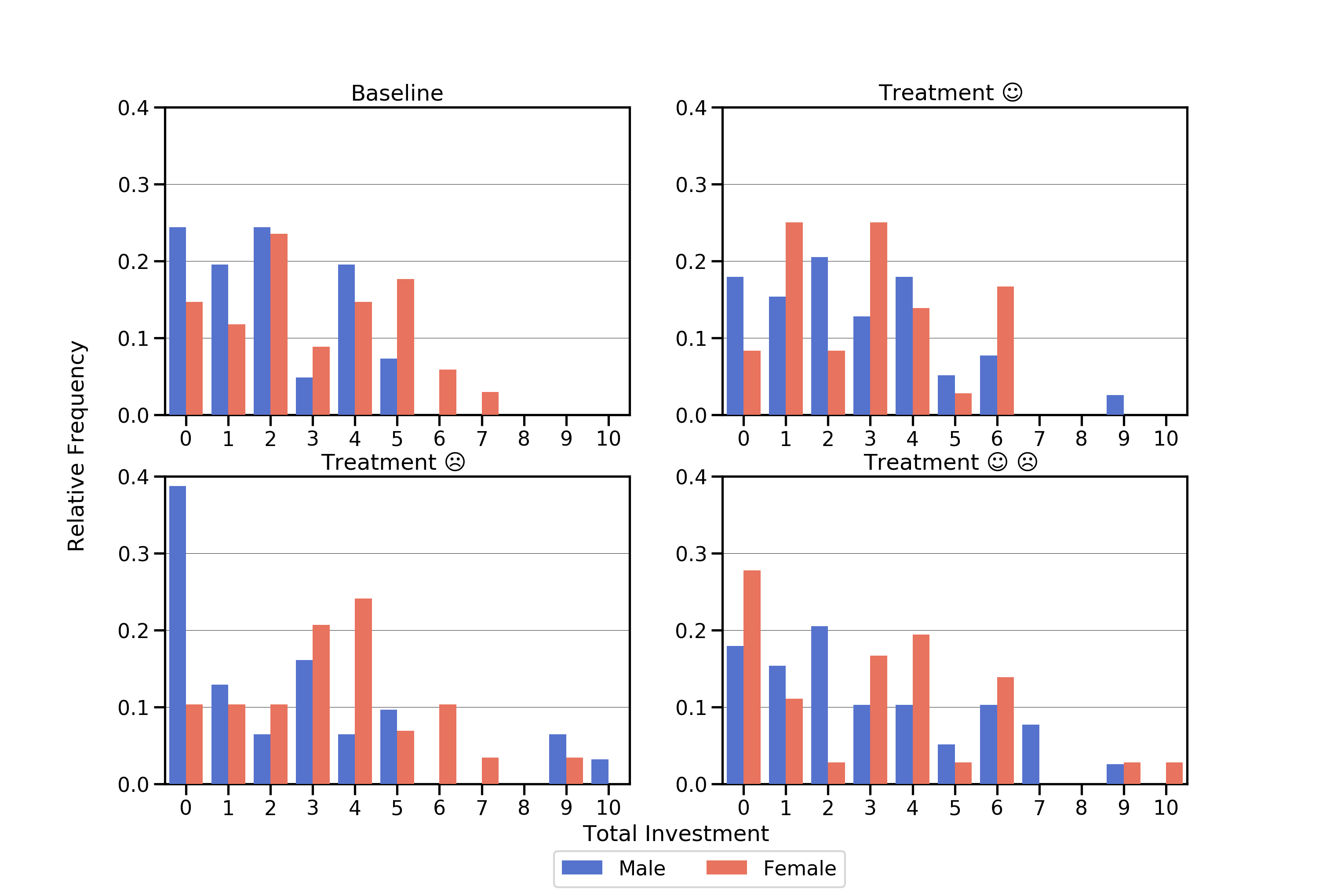}
        \caption{Distribution of Total Investment across Treatments}
        \label{fig:investment_gender_distribution}
\end{figure}

\begin{figure}[h!]    
\centering
        \includegraphics[width=\textwidth]{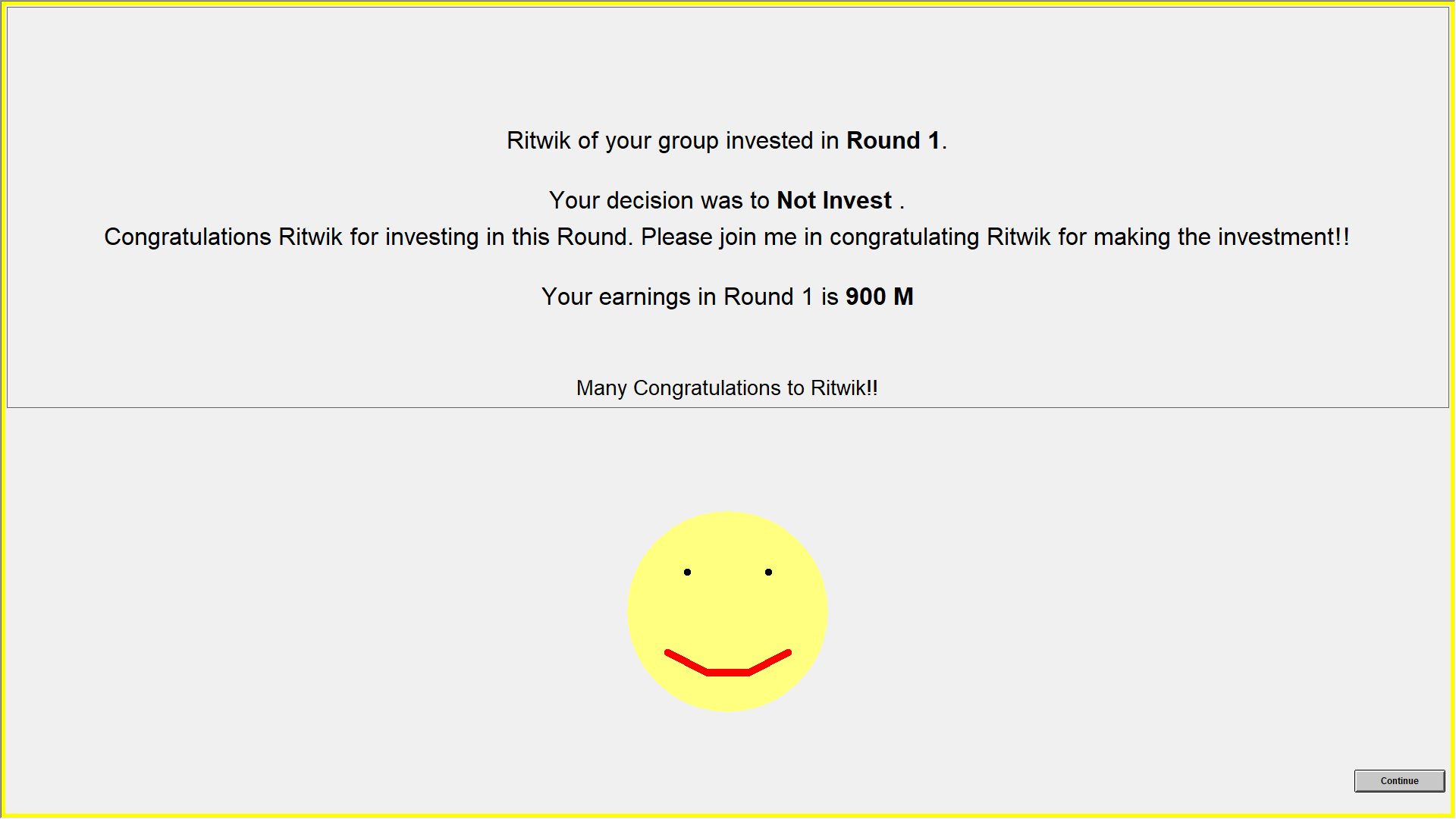}
        \caption{Screen with the Positive Social Recognition}
        \label{fig:psr}
\end{figure}

\begin{figure}[h!]    
\centering
        \includegraphics[width=\textwidth]{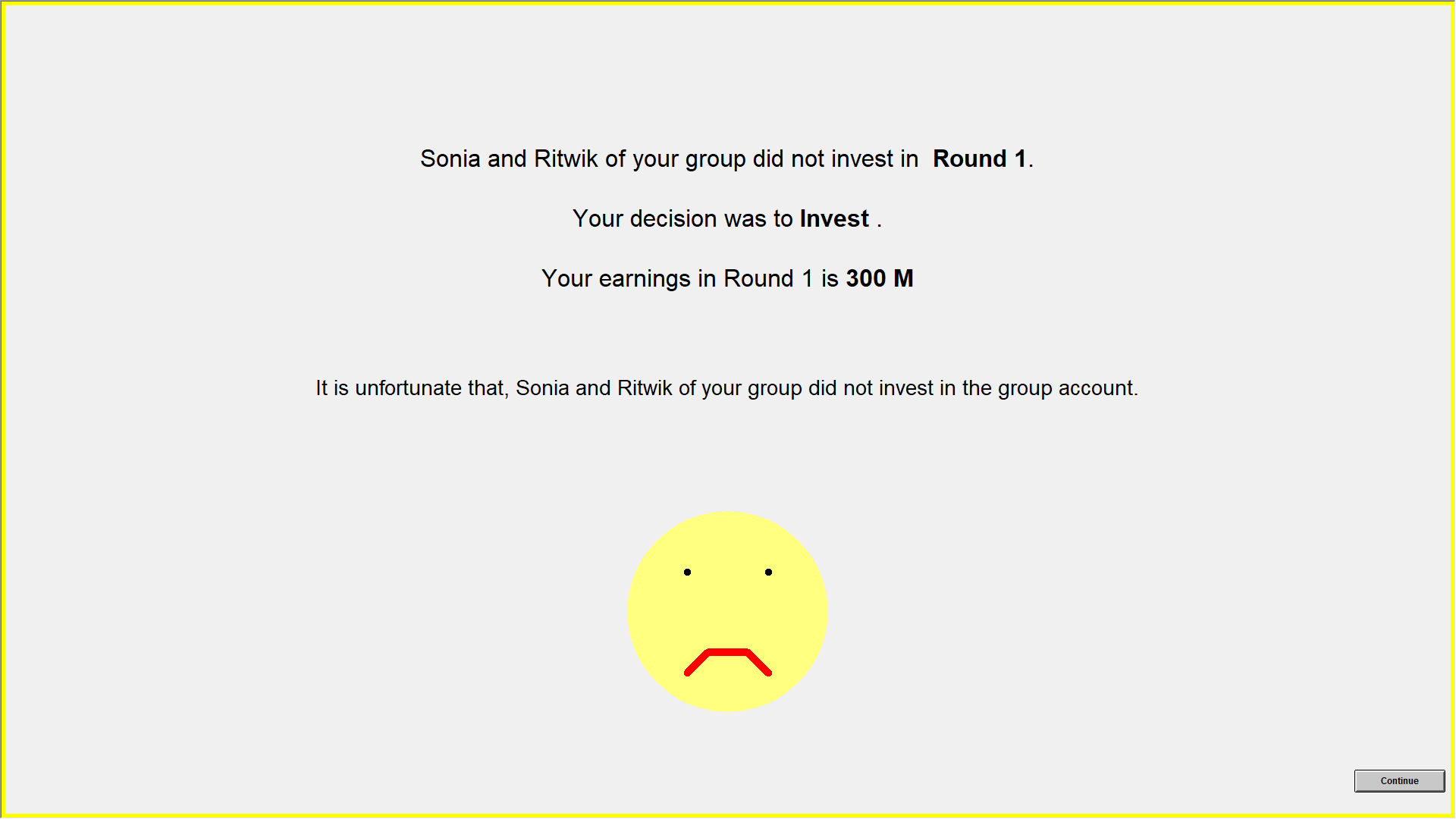}
        \caption{Screen with the Negative Social Recognition}
        \label{fig:nsr}
\end{figure}

\begin{figure}[h!]    
\centering
        \includegraphics[width=\textwidth]{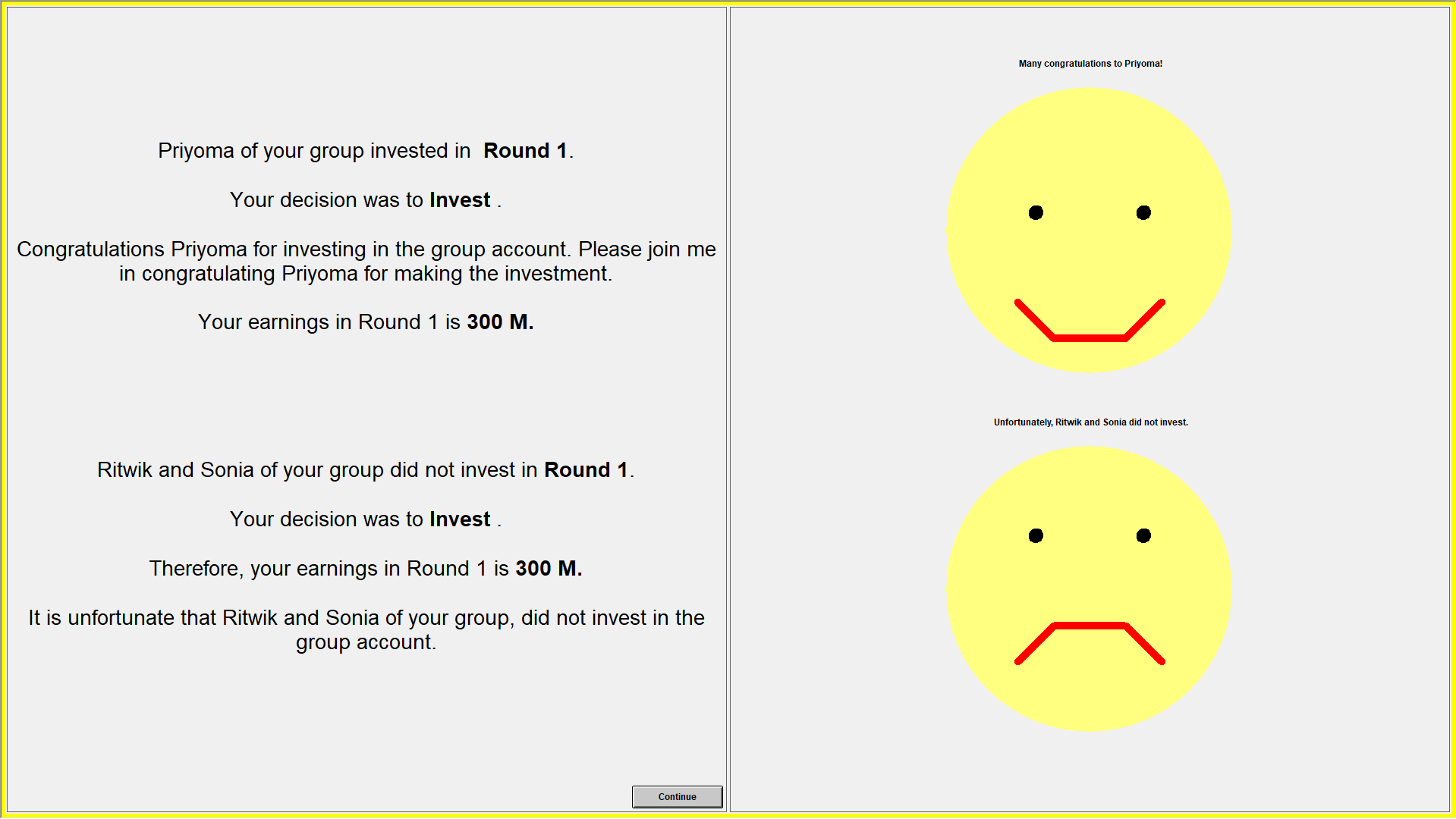}
        \caption{Screen with the Positive and Negative Social Recognition}
        \label{fig:pnsr}
\end{figure}
\clearpage
\section{Experimental Instructions}\label{Section-ExpIns}



\section*{Supplementary material}

# B  Experimental Instructions

# Experimental Instructions

## Informed Consent

Please read the sheet which has been handed over to you, carefully.

☐ I have understood what has been explained to me regarding the procedure and purpose of the research. I have understood my rights and offer my participation through a complete voluntary consent.

Name _____

Please enter your age _____

Please enter your gender
○ Male
○ Female

---

# Welcome

- Welcome to an experiment on decision making!
- Thank you for participating in our study. This is an experiment about economic decision making. The other people in this room are also participating in the experiment.
- You are free to leave the experiment at any time. However, if you choose to leave the experiment, you will have to forego your earnings.
- All decisions have to be private taken. You must not talk to the others or communicate with them in any way.
- If you have a question, please raise your hand and one of us will come to where you are sitting to answer it.

# Choose Names

As a first step, please choose a name for yourself from the list provided below. This is an anonymous experiment. As a result, all future references to another participant will be in terms of the chosen name. You will retain that name throughout the experiment. No two people can choose the same name. Suppose two people choose the same name, the first one to click 'submit' will be allowed to retain the name. The other person has to choose again.

**Choose your name:**
- Aditi
- Simran
- Kanika
- Tania
- Urvashi

- Tanmay
- Anup
- Sourav
- Rahul
- Udit

| | |
|---|---|
| Aditi | Tanmay |
| Simran | Anup |
| Kanika | Sourav |
| Tania | Rahul |
| Urvashi | Udit |

**10 names per group**

---

# Earnings

- For participating in the study today, you will earn at least **Rs. 50**. This is your participation fee for today.
- During the experiment, you will earn in terms of our laboratory currency called "Mohor" or **M** (meaning gold coins in ancient India). All Mohors you earn in the experiment will be converted into Rupees at the end of the experiment.
- The conversion rate is: 1 Mohor = Rs. 0.20
- The experiment will consist of **eleven rounds**. You will be paid for one **randomly chosen round between Round 1 and 10** in addition to your earning from **Round 11**.
- But you do not know which round you will be paid for, so play all the rounds carefully.

# Overview

- Here is a description of one of the first **ten rounds**.
- In each round you are randomly paired with two other participants to form a group. **You will never be paired with the same participant twice in a row.**
- The group remains the same through the round, but changes every round.
- Your earning in a given round depends on the decisions made by you and by your group members.
- Remember, you will be randomly paid for a round between Round 1 and 10.
- At the end of the experiment you will be paid, in private and in cash.

# Decision Stage Instructions

- In each round, you and the two other group members will have an opportunity to **invest** in a group account.
- You and your group members will have a time limit between **45- 90 seconds** to individually decide whether you want to invest in the group account.
- You will not get to know in advance the exact time limit you have for making the investment. For example, Group 1 may have a time limit of 48 seconds but Group 2 may have a time limit of 78 seconds.
- A sound clip (counting seconds in intervals of 5) will be played to help you understand how much time has passed.
- The round ends and the decision screen **disappears** as soon as you or a member of your group invests in the group account or when the stipulated seconds are up.

- Whichever group member invests first from the group, he/she will be will earn **300M** while the other two group members will earn **900M** in the first stage.
- If no investment is made by anyone in the group account within timeout, all three members of the group will earn **100M** for the first stage.
- If two group members simultaneously decide to invest, then it is **randomly** determined who earns 300M and 900M in the first stage.
- At the end of the round, the **chosen name** of the investor (if any) will be displayed on all three group members' screens.

**Treatment Positive SR** ☺

# Decision Stage Instructions

**All Subjects**

- In each round, you and the two other group members will have an opportunity to **invest** in a group account.
- You and your group members will have a time limit between **45- 90 seconds** to individually decide whether you want to invest in the group account.
- You will not get to know in advance the exact time limit you have for making the investment. For example, Group 1 may have a time limit of 48 seconds but Group 2 may have a time limit of 78 seconds.
- A sound clip (counting seconds in intervals of 5) will be played to help you understand how much time has passed.
- The round ends and the decision screen **disappears** as soon as you or a member of your group invests in the group account or when the stipulated seconds are up.
- Whichever group member invests first from the group, he/she will be will earn **300M** while the other two group members will earn **900M** in the first stage.
- If no investment is made by anyone in the group account within timeout, all three members of the group will earn **100M** for the first stage.
- If two group members simultaneously decide to invest, then it is **randomly** determined who earns 300M and 900M in the first stage.
- At the end of the round, the **real name** of the investor (if any) will be displayed on all three group members' screens.

---

**Treatment Negative SR** ☹

# Decision Stage Instructions

**All Subjects**

- In each round, you and the two other group members will have an opportunity to **invest** in a group account.
- You and your group members will have a time limit between **45- 90 seconds** to individually decide whether you want to invest in the group account.
- You will not get to know in advance the exact time limit you have for making the investment. For example, Group 1 may have a time limit of 48 seconds but Group 2 may have a time limit of 78 seconds.
- A sound clip (counting seconds in intervals of 5) will be played to help you understand how much time has passed.
- The round ends and the decision screen **disappears** as soon as you or a member of your group invests in the group account or when the stipulated seconds are up.
- Whichever group member invests first from the group, he/she will be will earn **300M** while the other two group members will earn **900M** in the first stage.
- If no investment is made by anyone in the group account within timeout, all three members of the group will earn **100M** for the first stage.
- If two group members simultaneously decide to invest, then it is **randomly** determined who earns 300M and 900M in the first stage.
- At the end of the round, the **real name** of the two non-investors will be displayed on all three group members' screens.

`Treatment Positive Negative SR ☺ ☹`  `All Subjects`

# Decision Stage Instructions

- In each round, you and the two other group members will have an opportunity to **invest** in a group account.
- You and your group members will have a time limit between **45- 90 seconds** to individually decide whether you want to invest in the group account.
- You will not get to know in advance the exact time limit you have for making the investment. For example, Group 1 may have a time limit of 48 seconds but Group 2 may have a time limit of 78 seconds.
- A sound clip (counting seconds in intervals of 5) will be played to help you understand how much time has passed.
- The round ends and the decision screen **disappears** as soon as you or a member of your group invests in the group account or when the stipulated seconds are up.

- Whichever group member invests first from the group, he/she will be will earn **300M** while the other two group members will earn **900M** in the first stage.
- If no investment is made by anyone in the group account within timeout, all three members of the group will earn **100M** for the first stage.
- If two group members simultaneously decide to invest, then it is **randomly** determined who earns 300M and 900M in the first stage.
- At the end of the round, the **real name** of the investor and **real name** of two non-investor will be displayed on all three group members' screens.

`Common to all Treatments`  `All Subjects`

# Quiz for Instructions

- Will you play with the same group members in every round? [Yes, No]
- How much time do you have to decide? [45 secs, 90 secs, randomly pre-determined seconds between the range of 45-90 seconds.]
- Suppose in a round, you do not invest but someone else invests from your group. How much do you earn? [100M, 300M, 900M, NOTA]
- Suppose in a round, you invest before anyone else in the group. How much do you earn? [100M, 300M, 900M, NOTA]
- Suppose in a round, no one from your group invests. How much do you earn? [100M, 300M, 900M, NOTA]
- Which round will you be paid for? [All, random from 1-10 & 11, Last, First]

# Please Begin

The decision rounds will now begin.

Remember the following:

1. There are **ten rounds** of decision making.
2. The time limit for each round will be randomly picked from 45sec and 90sec.
3. In each of the following rounds you are randomly matched with two other people in this room.
3. You are never matched with the same person in two consecutive rounds.

# Decision Stage

- The red button in the centre of the screen is used to make your investment decision.
- The round ends and the decision screen disappears as soon as you or a member of your group invests in the group account.
- Please click this button if you wish to invest.

**INVEST**

You have been randomly paired with two participants. If no member of your group invests then you will each make 100M. If a member of your group invests then that member will make 300M, and the other two group members will each make 900M.

| Common to all Treatments | **Round 1-10** | **All 3 Subjects in each group** |

## Screens after round ends

**If no investor in the group**

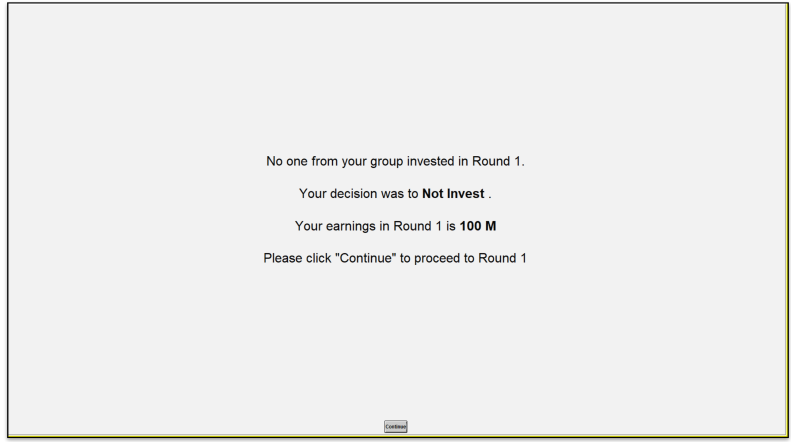

| Treatment Baseline | **Round 1-10** | **All 3 Subjects in each group** |

## Screens after round ends

**At least one investor in the group**

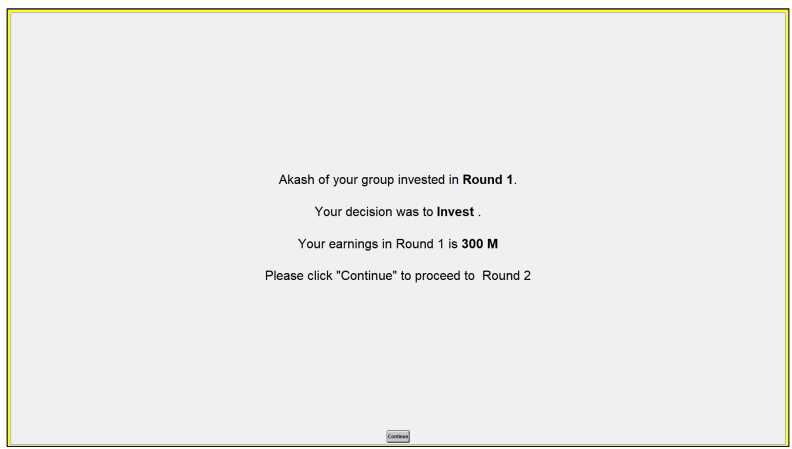

**Treatment Positive SR** ☺          **Round 1-10**          **All 3 Subjects in each group**

# Screens after round ends

**At least one investor in the group**

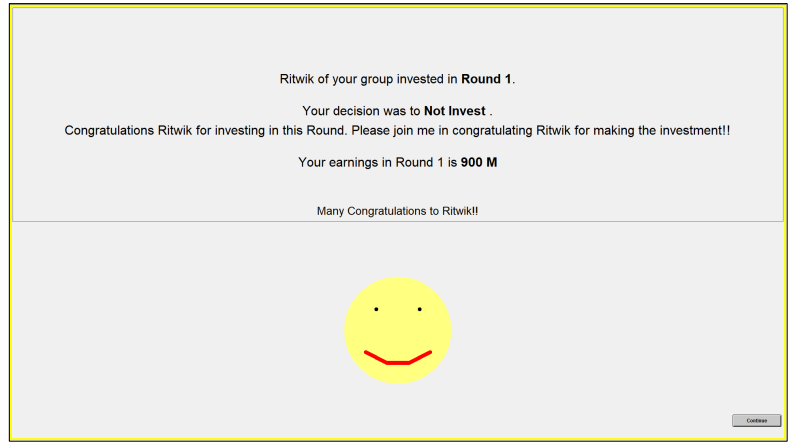

**Treatment Negative SR** ☹          **Round 1-10**          **All 3 Subjects in each group**

# Screens after round ends

**At least one investor in the group**

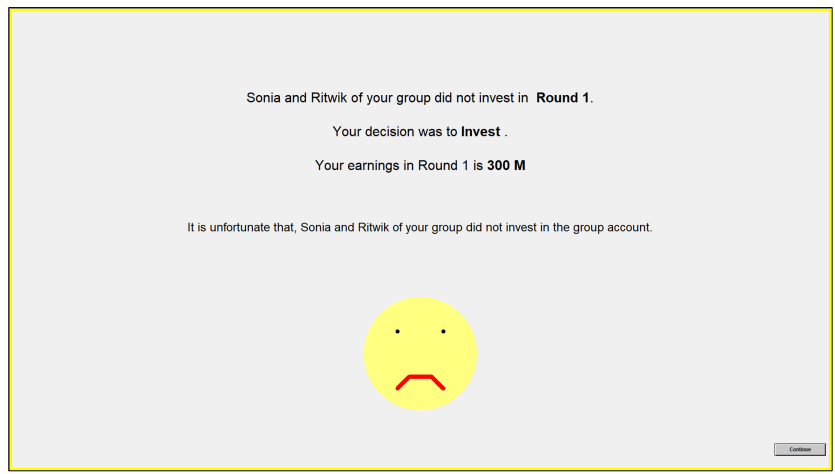

| Treatment Positive Negative SR ☺ ☹ | **Round 1-10** | **All 3 Subjects in each group** |

# Screens after round ends

### At least one investor in the group

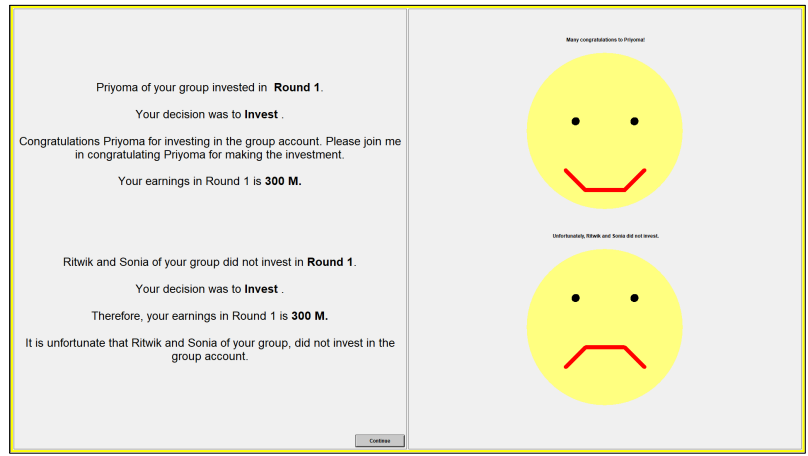

| Common to all Treatments | **Round 11** | **All Subjects** |

# Risk

- Your decision sheet shows ten decisions listed on the left. Each decision is a paired choice between **"Option A"** and **"Option B."**
- You will make **ten choic**es and record these in the final column, but only one of them will be randomly chosen by the computer in the end to determine your earnings.
- Even though you will make ten decisions, **only one** of these will end up affecting your earnings for Round 11, but you will not know in advance which decision will be used. Obviously, each decision has an equal chance of being used in the end.
- Now, please look at Decision 1 at the top. Option A pays 200M with 10% chance, and it pays 160M with 90% chance. Option B yields 385M with 10% chance, and it pays 10M 90% chance. The computer program will determine the chances and calculate accordingly.
- The other Decisions are similar, except that as you move down the table, the chances of the higher payoff for each option increase. In fact, for Decision 10 in the bottom row, the random decision will not be needed since each option pays the highest payoff for sure, so your choice here is between 200M or 385M.
- So now please look at the empty boxes on the right side of the record sheet. You will have to record your decision between A or B in each of these boxes.

# Risk

- Out of the ten choices you made the randomly chosen choice is <>
- You had chosen Option <x> for this choice.
- Now the computer will choose <>M with <>% chance and <>M with <>% chance.
- You have won <>M because <>M was randomly chosen by the computer.
- Your earning for Round 11 is <>M

# Behavioral and Personality Traits:
**(Answers to all question was presented in on a scale 1-5 , with radio buttons)**

1. If you win a prize for the best all round performance in your university, how likely are you to inform your friends about it?

2. If you win a prize for the best all round performance in your university, how likely are you to inform your family about it?

3. If you win a prize for the best all round performance in your university, how likely are you to post on Facebook or other social media about it

4. How stressed did the time make you feel during the decision task?

5. How do you see yourself: are you generally a person who is fully prepared to take risks or do you try to avoid taking risks?

6. How likely are you to admit that your tastes are different from those of your friends?

7. How likely are you to argue with a friend about an issue on which he or she has a very different opinion?

8. How likely are you to defend an unpopular issue that you believe in at a social occasion?

9. People should be willing to help others who are less fortunate

10. These days people need to look after themselves and not overly worry about others.

11. It is important to help one another so that the community in general is a better place.

12. I see myself as someone who tends to find fault with others

13. I see myself as someone who can be cold and aloof

14. I see myself as someone who is considerate and kind to almost everyone

15. I see myself as someone who likes to cooperate with others

16. I see myself as someone who is sometimes rude to others

17. I see myself as someone who is helpful and unselfish with others

18. I see myself as someone who starts quarrels with others

19. I see myself as someone who has a forgiving nature

20. I see myself as someone who is generally trusting

# Demographics:

Gender:
- Male
- Female

Caste:
- General
- Scheduled Caste
- Scheduled Tribe
- OBC
- Prefer not to say

Religion:
- Hindu
- Muslim
- Christian
- Prefer not to say

Which year in college are you?

- UG First Year
- UG Second Year
- UG Third Year
- UG Fourth Year
- Masters First Year
- Masters Second Year

What is your approximate monthly family income in Rupees?

_____

_____

(Free form question at the end)

We are interested in the reasons why you did or did not choose to invest in this study. Please explain what affected your decisions.

_____



%\begin{itemize}
 
  %  \item \scriptsize In Baseline, the per round decrease in the investment rate is significant.(Marginal effect -0.029 , p = 0.003). For treatment \smiley{} , \frownie{} and \smiley{}\frownie{} the per round increase in the investment rate is not significant.(Marginal effect 0.003 , p = 0.671 ; (Marginal effect -0.003 , p = 0.686 respectively ; Marginal effect -0.012 , p = 0.088)  
  %  \item \tiny \textcolor{blue}{ \italic{Since figure c talks shows the distribution , maybe its better to talk about ranksum , ksmirnov(especially) and fisher's exact by gender within each treatment here : these are only significant for pos+neg}}
%\end{itemize}